\documentclass[%
 reprint,
 amsmath,amssymb,
 aps,
 pra,
 floatfix, 
]{revtex4-2}

\usepackage{graphicx}
\usepackage{dcolumn}
\usepackage{bm}
\usepackage{comment}
\usepackage{amsmath}
\usepackage{parskip}
\usepackage[utf8]{inputenc}
\usepackage[T2A]{fontenc}
\DeclareUnicodeCharacter{2010}{-}

\begin{document}

\title{Unusual sign-changing Faraday effect in nanometer-thick magnetic films}

\author{Belkova A.V.}
 \email{belkova.av18@physics.msu.ru}
 \affiliation{Faculty of Physics, Lomonosov Moscow State University, Leninskie gori, Moscow 119991, Russia}
\author{Ignatyeva D.O.}
 \affiliation{Faculty of Physics, Lomonosov Moscow State University, Leninskie gori, Moscow 119991, Russia}
\affiliation{Russian Quantum Center, Moscow 121205, Russia}
\author{Kalish A.N.}
 \affiliation{Faculty of Physics, Lomonosov Moscow State University, Leninskie gori, Moscow 119991, Russia}
\affiliation  {Russian Quantum Center, Moscow 121205, Russia}
\author{Vetoshko P.M.}
\affiliation{Institute of Physics and Technology, V.I. Vernadsky Crimean Federal University, Simferopol 295007, Crimea}
\author{Kudryashov A.L.}
\affiliation{Institute of Physics and Technology, V.I. Vernadsky Crimean Federal University, Simferopol 295007, Crimea}
\author{Belotelov V.I.}
 \email{belotelov@physics.msu.ru}
\affiliation{Faculty of Physics, Lomonosov Moscow State University, Leninskie gori, Moscow 119991, Russia}
\affiliation{Russian Quantum Center, Moscow 121205, Russia}

\date{\today}

\begin{abstract}
It is generally believed that the magneto-optical Faraday effect appears in the the bulk of a magnetic material and its sign is fully determined by the sign of the non-diagonal permittivity element. Here we reveal an additional contribution to the Faraday effect from the film interfaces. It becomes notable for films with a thickness of a few tens of nanometers. As a result, in the absorption band of the film a novel feature of the Faraday effect is experimentally observed and numerically confirmed: sign of the Faraday rotation at a fixed wavelength becomes dependent on the film thickness and therefore is ambiguously related to the sign of the gyration and magnetization of the film. We elaborated an analytical model taking into account an interplay between the bulk and surface contributions which nicely describes the experimental data. Moreover, the Faraday rotation coming purely from interfaces without any bulk contribution is demonstrated. Possible applications of the observed unusual Faraday rotation behavior include hardware data encryption and other devices performing polarization control at nanoscale.
\end{abstract}
\maketitle
The magneto-optical Faraday effect, which is a rotation of the polarization plane of light passing through a magnetized material along its magnetization, was discovered by Michael Faraday in 1845~\cite{faraday1846magnetization}. Since then, numerous theoretical and experimental studies were performed to investigate this effect in detail (see~\cite{inoue2013magnetophotonics} and references therein) both in smooth films~\cite{deb2012magneto,levy2015large,PhysRevLett.130.166901},
and nanostructures~\cite{lyubchanskii2003magnetic,PhysRevApplied.22.044064,elsayed2025role}.

Nowadays, the Faraday effect has a great practical significance for various nonreciprocal devices, such as isolators~\cite{srinivasan2022review,karki2019toward}, circulators~\cite{smigaj2010magneto}, modulators~\cite{silva2015assembly}, rotators~\cite{li2020bridgman}, deflectors, routers~\cite{ho2018switchable}, sensors~\cite{maccaferri2015ultrasensitive,chen2023review} and magnetometers~\cite{ignatyeva2021vector}, data encryption~\cite{bi2022magnetically}  and other devices~\cite{satoh2012directional}. Currently, for the purpose of device miniaturization and integration the Faraday effect in nanometer thick films is of prime importance ~\cite{kulikova2023faraday}.

\begin{figure*}[htb]
\begin{minipage}[h]{0.49\linewidth}
 \begin{flushleft}
    (a)
  \end{flushleft}
\center{
\includegraphics[width=1\linewidth]{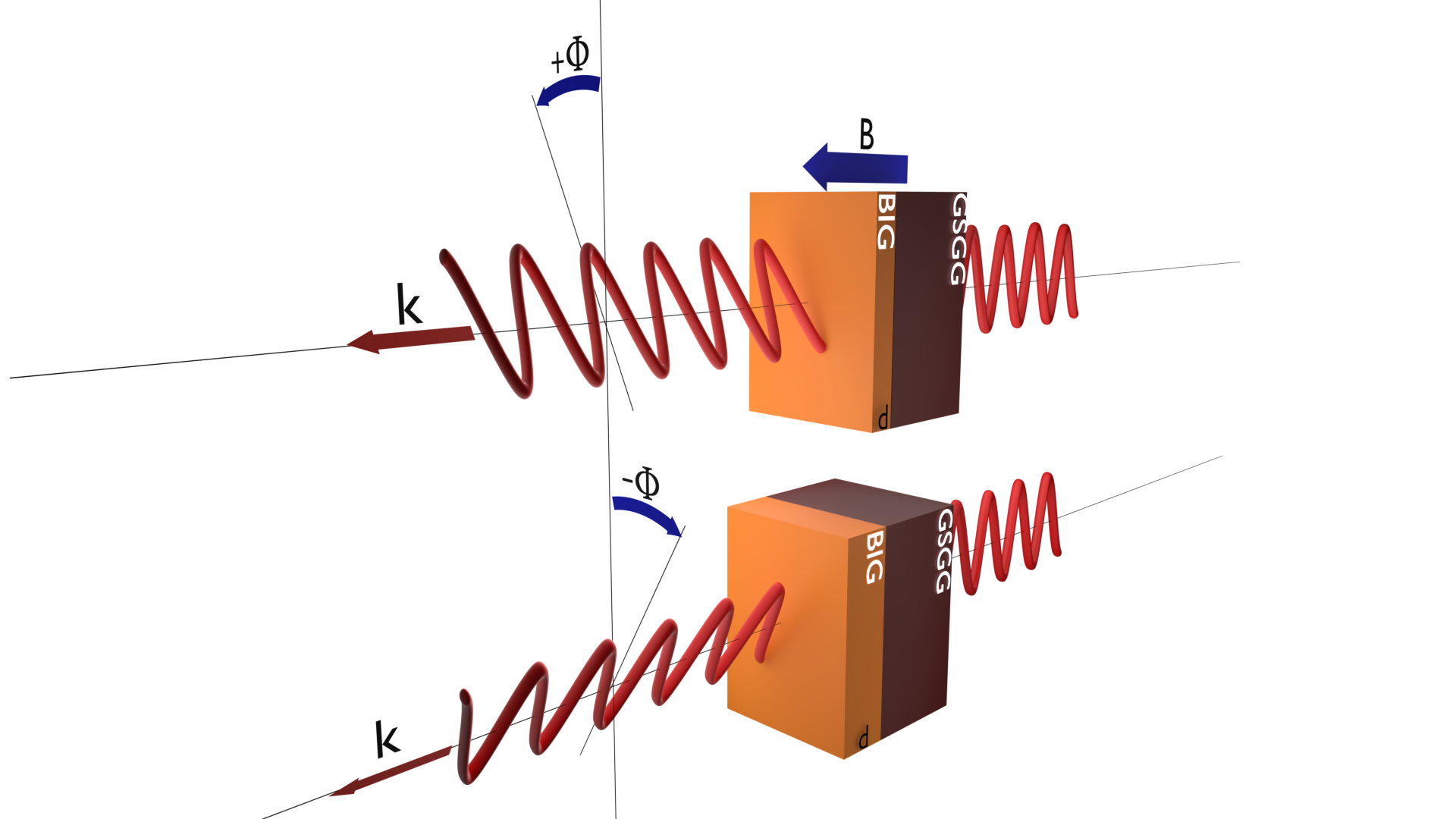}}\\
 \begin{flushleft}
    (b)
  \end{flushleft}
\center{\includegraphics[width=1\linewidth]{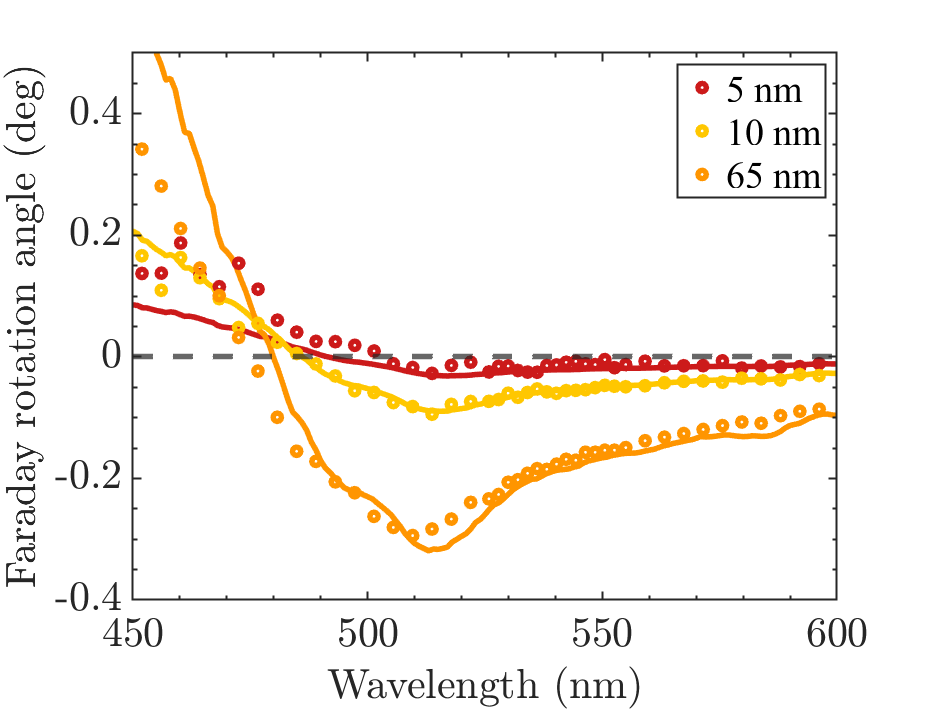}} \\
\end{minipage}
\hfill
\begin{minipage}[h]{0.49\linewidth}
\begin{flushleft}
    (c)
  \end{flushleft}
\center{\includegraphics[width=1\linewidth]{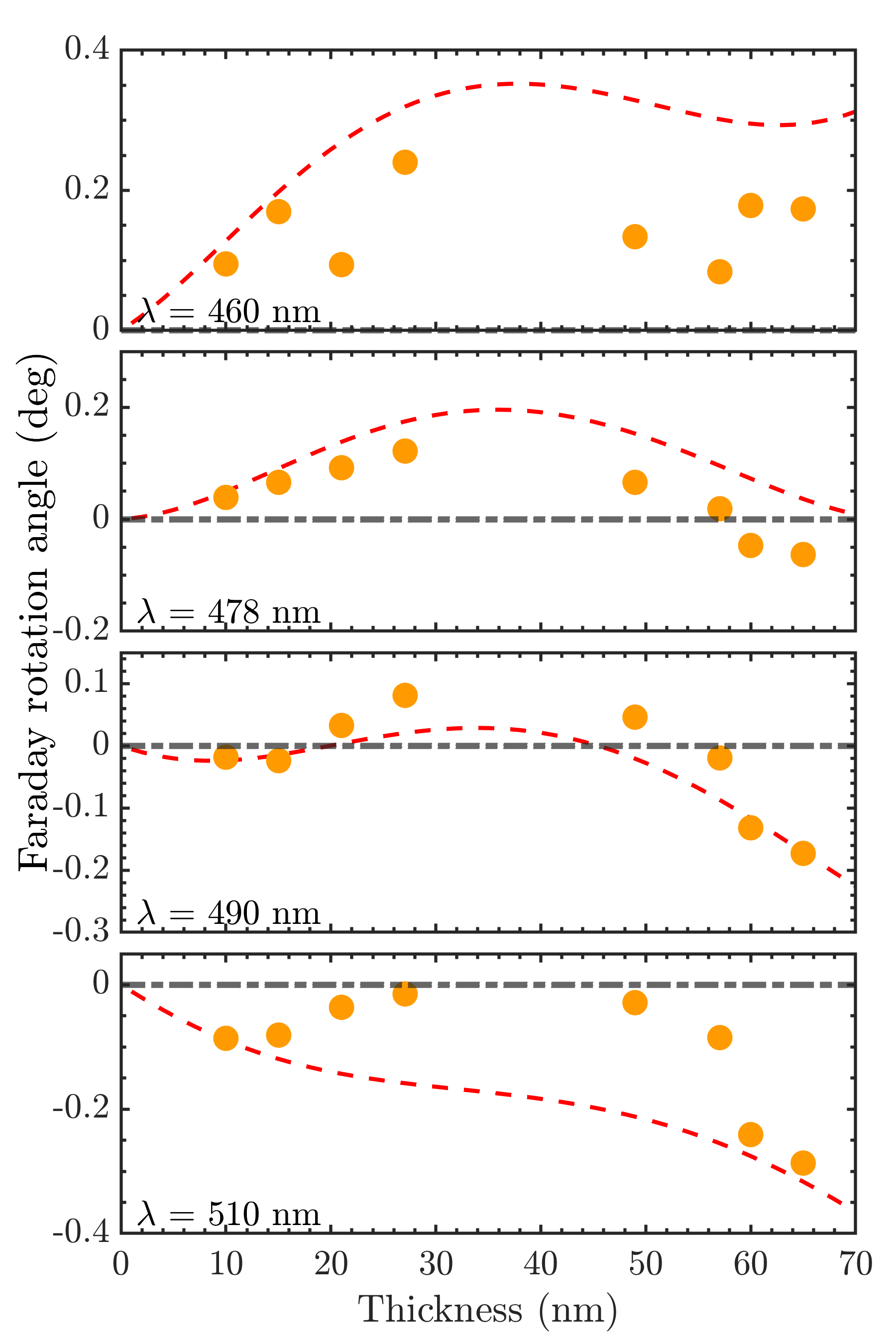}} \\
\end{minipage}
\caption{\label{Fig: General}(a) Scheme of the configuration with opposite FR for the two different thicknesses of magnetic film illuminated by light at a fixed wavelength. (b) The experimentally obtained FR spectra for three films of thicknesses $d$=5~nm, $d=$10~nm, and $d=65$~nm (open circles) compared with calculated spectra (solid curves). (c) Dependencies of the FR angle on the film thickness at light wavelength of 465~nm (first panel), 478~nm (second panel), 490~nm (third panel) and 510~nm (fourth panel), experimentally measured (orange circles) and calculated (dashed lines).}

\end{figure*}

Let's consider a homogeneous isotropic magnetic film of thickness $d$ described by the dielectric permittivity tensor $\varepsilon_{ij}=n_m^2\delta_{ij}+ie_{ijk}g_k$, where $\delta_{ij}$ is the Kronecker delta, $i$ is the imaginary unit and $e_{ijk}$ is the Levi-Civita symbol. Diagonal elements of the dielectric tensor are given by the complex refractive index $n_m=n_m'+in_m''$, and non-diagonal elements are expressed by the gyration vector $g$, which is also complex valued: $g=g'+ig''$. Therefore, we deal with a magnetic film at the wavelength range inside the absorption band. In this case the Faraday rotation (FR) angle $\Phi$ is given by~\cite{zvezdin1997modern}:
\begin{equation}
    \Phi = \frac{\pi}{\lambda}d\frac{g'n_m' + g''n_m''}{|n_m^{2}|},
\label{Eq: Far from book}
\end{equation}
which implies that for a fixed wavelength $\lambda$ the sign of $\Phi$ is determined by the term $(g'n_m' + g''n_m'')$. It agrees with experimental studies which demonstrate that, in contrast to some complicated cases, such as anisotropic crystals~\cite{kurtzig1969magneto,krinchik1969transparent,ignatyeva2022birefringence} and nanostructured materials~\cite{belotelov2006magnetooptics,armelles2008localized,floess2017plasmonic,kuzmin2016giant,christofi2018giant,royer2020enhancement,almpanis2016metal,xia2022enhancement}, for isotropic homogeneous films the sign of the FR at a fixed wavelength is the same for any film thickness, although the FR angle value might depend on the film thickness in a non-monotonic way~\cite{levy2019faraday}. 

Surprisingly, we found that in smooth homogeneous isotropic nanometer-thick films the FR angle behavior is in contradiction with Eq.~\eqref{Eq: Far from book}, and modification of the film thickness changes not only the value, but also the sign of the FR angle. We experimentally and theoretically demonstrate these new sign-changing FR behavior at a fixed wavelength in the vicinity of the magneto-optical resonance. We elaborate an analytical model to explain origin of such unusual properties. It should be noted that in a sharp contrast to anisotropic crystals, where the sign-changing behavior is caused by the linear birefringence and requires dozens of micrometers to reveal, the sign-change properties discussed here are observed at nanometer scales. This opens unique opportunities for practical applications, and we show that such a fundamental effect might be used, for example, for the hardware data encryption.

We perform experimental studies of the Faraday effect in a series of nanometer-thick bismuth-substituted iron-garnet films of $\mathrm{Bi_1Gd_2Fe_{4.4}Sc_{0.6}O_{12}}$  composition. Initially, a 200~nm thick film was grown on gadolinium-scandium gallium garnet (GSGG) substrate by liquid-phase epitaxy. Further, sections of thicknesses ranging from 5~nm to 65~nm were obtained by the method of inhomogeneous lateral reactive ion etching of this film.

Experimental measurements of the FR were carried out using the balanced scheme (Supplementary C) with the studied samples placed in the external magnetic field of $\pm200$~mT as schematically shown in Fig.~\ref{Fig: General}(a). 
\begin{figure*}[htbp!]
    \includegraphics[width=1\textwidth]{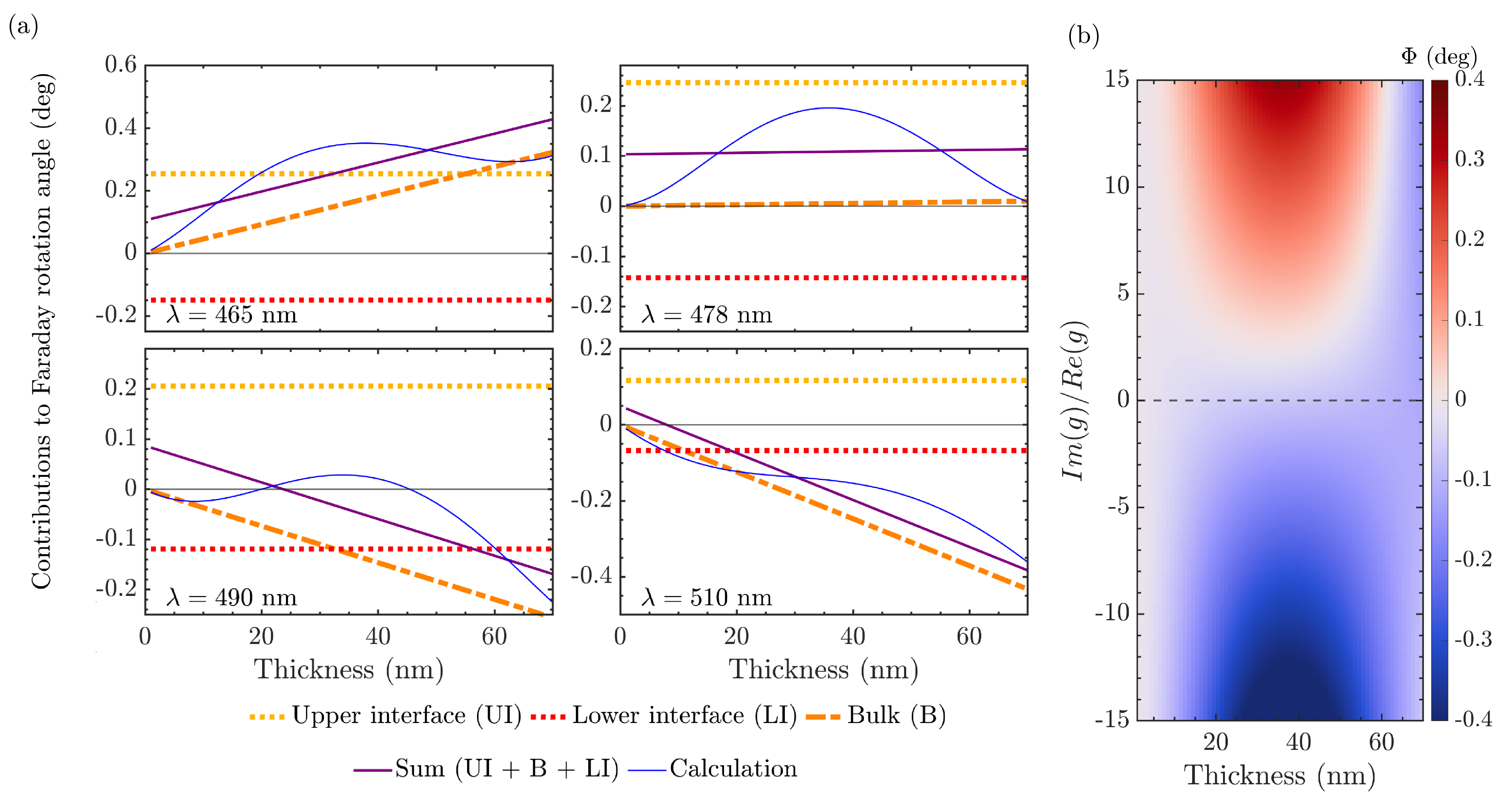}
\caption{\label{Fig: Contributions} Physical origin of the sign-changing FR. (a) Interface and bulk contributions (see legends below the graphs) to the FR as a function of the film thickness for the wavelengths $\lambda=465$~nm, $\lambda=478$~nm, $\lambda=490$~nm, $\lambda=510$~nm. (b) The FR dependence on the film thickness $d$ and ratio of imaginary and real parts of the gyration $g''_m/g'_m$. The real part of $g$ is $-0.013$, $\varepsilon = 7.16 + 0.65i$. Wavelength is  $\lambda=480$~nm.}
\end{figure*}
Comparing the FR spectra measured for three samples of different thicknesses (Fig.~\ref{Fig: General}(b), open circles), one might notice a remarkable and surprising feature: the Faraday angle turns to zero at different wavelengths ($\lambda_{0}=$505~nm for $d=5$~nm, $\lambda_{0}=$485~nm for $d=10$~nm, and $\lambda_{0}=$474~nm for $d=65$~nm). It leads to the sign-changing behavior of the FR for varying film thicknesses at some fixed wavelengths. Thus at $\lambda=$478~nm the FR angle is positive for the films of thicknesses up to 70~nm and then turns negative (Fig.~\ref{Fig: General}(c), second panel, Supplementary Fig.~S3(a), second panel). At a bit larger wavelength of $\lambda=$490~nm the FR sign change takes place at around thickness of 20~nm (Fig.~\ref{Fig: General}(c), third panel). Moreover, additional sign change appears at 55~nm, so that the FR becomes negative again for the thicker films. At the same time,  at some neighbor wavelengths, e.g. at 465~nm and  510~nm FR doesn't change sign and keeps either positive or negative for all thicknesses, respectively (Fig.~\ref{Fig: General}(c), first and fourth panels).

The observed experimental results agree with theoretical calculations  performed based on the exact analytical expression defining the FR angle $\Phi$ as the half-sum of the phase $\phi_\pm$ difference acquired by the two circularly polarized waves of opposite helicities during their propagation through a magnetic medium~\cite{zvezdin1997modern}: $\Phi=\frac{1}{2}(\phi_{+}-\phi_{-})$, where '$+$' and '$-$' correspond to the two circular polarizations. The phases $\phi_\pm$ are determined by the arguments of the complex amplitude transmission coefficients and the resulting FR angle is calculated by Eq.~(S.5) from Supplementary B.  

Using the same diagonal and off-diagonal components of the magnetic film dielectric tensor for all of the films (Supplementary, Fig.~S2) we obtain a good agreement of the calculated spectra by Eq.~(S.5) with the experimentally observed ones (compare solid curves for calculated FR with dots demonstrating measured FR in Fig.~\ref{Fig: General}(b)). Therefore, the observed unusual behavior of FR is confirmed by calculations.   
\begin{figure*}[htbp!]
    \includegraphics[width=0.8\linewidth]{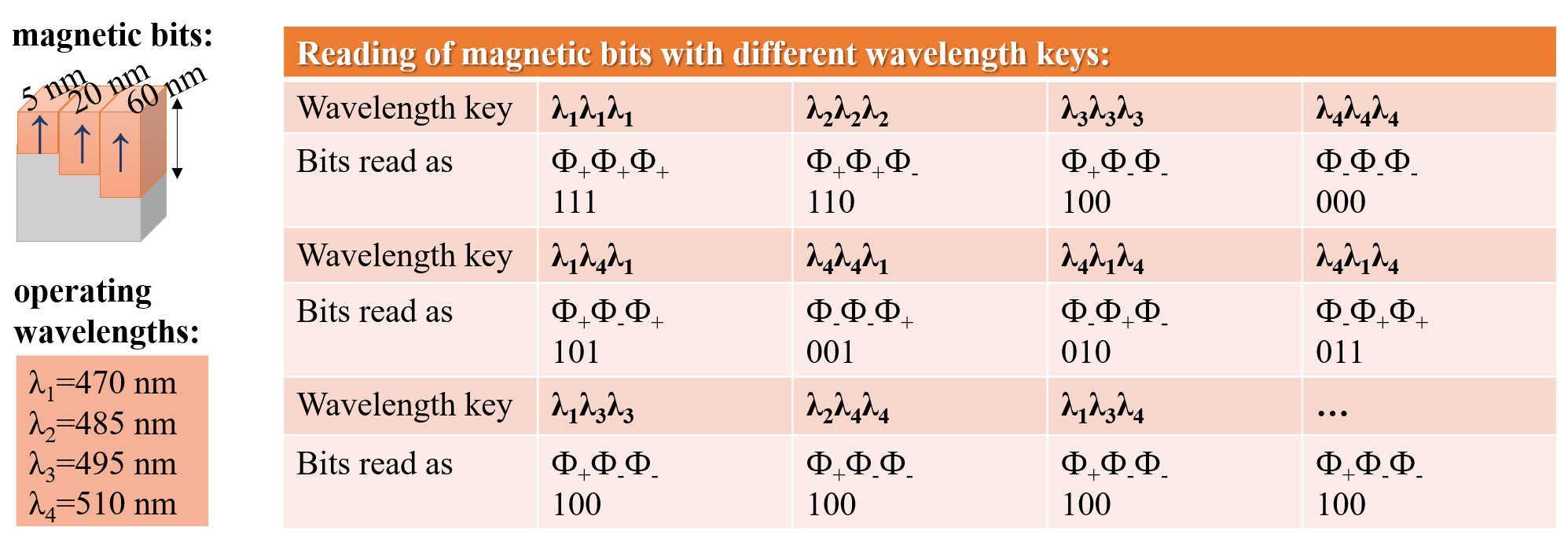}
    \caption{\label{Fig: Encryption}Schematic example of data encryption showing 3 bits with 'up' magnetization read with different combinations of the wavelength keys.}
\end{figure*}

To explain the origin of the observed phenomenon let's start from analyzing light propagation using a simple, but rather illustrative model that takes into account only a single pass of the light through the film and neglects all of the interference effects. The crucial point is that to determine the phase difference acquired by the two circularly polarized waves of opposite helicity, one should take into account not only the contribution of the bulk material $\Delta \phi_\mathrm{bulk}$, as is usually considered, but also the phase difference acquired by light passing through the upper ($\Delta \phi_\mathrm{up}$) and lower ($\Delta \phi_\mathrm{low}$) interfaces of the magnetic film:
\begin{equation} \label{eq:2}
\begin{split}
\Phi & = \frac{1}{2}(\Delta\phi_\mathrm{up}+\Delta\phi_\mathrm{bulk}+\Delta\phi_\mathrm{low}) \\
\Delta \phi_\mathrm{up} & = \mathrm{arg}(t_{12+}) - \mathrm{arg}(t_{12-}) \\
\Delta \phi_\mathrm{bulk} & = \mathrm{Re}\left[\frac{g}{n_m}\right] k_0 d \\
\Delta \phi_\mathrm{low} & = \mathrm{arg}(t_{23+}) - \mathrm{arg}(t_{23-}),
\end{split}
\end{equation}
where $t_{ij\pm}=2 n_i / (n_i + n_j)$ are the complex amplitude transmission coefficients through the interface between media '$i$' and '$j$' for the right ('$+$') or left ('$-$') circular polarizations; indices $1,2,3$ denote the upper medium (air), magnetic material and lower medium (substrate), correspondingly; $n_{2\pm}=\sqrt{n_m^2\pm g}$, and $k_0$ is the vacuum wavenumber. 

We then assume that $n_m' \gg n_m'' \gg g'\sim g''$, which is true for the experimental samples (Supplementary, Fig.~S2). The FR angle within this single-pass model for a thin film, therefore, takes the form:

\begin{equation}
   \Phi \approx-\frac{g''(n_m'^2 - n_1n_3)}{2n_m'^2(n_1+n_m')(n_3+n_m')} +\pi\frac{d}{\lambda}\mathrm{Re}\left[\frac{g}{n_m}\right]
\label{Eq: FR simple}
\end{equation}

Obviously, the first term does not depend on the film thickness, while the second one is linear in it. If the signs of the first term and $\mathrm{Re}[\frac{g}{n_m}]$ are different, then the sign of the FR changes for a certain thickness $d$. Thus, the single-pass model provides an analytical explanation of the observed effect. Moreover, it gives an insight on the conditions when such behavior might be observed: (1) the imaginary part of the gyration $g''$ should be large enough (Fig.~\ref{Fig: Contributions}(b)), (2) the surface contributions to the FR should be comparable with the bulk one, which means that for $g''\sim g'$ the effect of FR sign change is observed at the thicknesses of $d\sim10-30$~nm.

Fig.~\ref{Fig: Contributions}(a) shows that, indeed, at $\lambda=490$~nm the sum of three contributions (purple line): the two surface ones $\Delta\phi_\mathrm{up}$ (horizontal dashed orange line), $\Delta\phi_\mathrm{down}$ (horizontal dashed red line) and bulk one $\Delta\phi_\mathrm{bulk}$(dash-dot orange line), changes the sign at $d=25$~nm which agrees with experiment. It should be noted that the simple model nicely follows curve calculated by Eq.~(S.5) (blue solid line). The exact solution oscillates with respect to it. The oscillations are due to the interference effect which is not taken into account within the single-pass model.

Quite unusual situation appears at $\lambda=478$~nm. Indeed, the bulk contribution vanishes since in accordance to Eq.~\eqref{Eq: FR simple} it is proportional to $\mathrm{Re}\left[\frac{g}{n_m}\right]$ which becomes zero (see horizontal dash-dot orange line in Fig.~\ref{Fig: Contributions}(a) for $\lambda=478$~nm). Therefore, at this wavelength only interfacial contribution to FR is present and the single-pass model predicts the FR angle independent on the film thickness. Such unusual case really takes place at $\lambda=478$~nm, however, due to the interference, the FR angle oscillates around the horizontal line for the total contribution of the interfaces (see blue solid line and horizontal purple line in Fig.~\ref{Fig: Contributions}(a) for $\lambda=478$~nm). It nicely agrees with experiment (Fig.~\ref{Fig: General}(c), second panel).

In contrast, the sum of all three contributions is positive at $\lambda=465$~nm  and no sign change is observed. At $\lambda=510$~nm the single-pass model predicts the sign change for $d=8$~nm but it is is completely undiminished by the interference oscillations which makes the FR negative for any film thickness (Fig.~\ref{Fig: Contributions}(a)). 

For more precise analysis one should take into account the interference effects in the film. Nevertheless, their contribution is just an additional oscillatory in $d$ term in Eq.~\eqref{Eq: FR simple} (Supplementary, Fig.~S3(a)), so they don't add anything significant to the origin of the effect. Therefore, the single-pass model nicely explains the physics of the sign-changing FR behavior. 

It is important to quantify the conditions under which the dependence of the FR on the thickness of the magnetic film is sign-changing. Fig.~\ref{Fig: Contributions}(b) shows that as predicted by Eq.~\eqref{Eq: FR simple}, high imaginary part of the gyration is required. The sign of the FR angle changes with thickness if the ratio $g''_m/g'_m > 3$ or $g''_m/g'_m < -10$.

Besides the certain fundamental interest, the discussed effect has a practical significance. For example, it might be used for the hardware stored data encryption. 
In the case of classical magnetic recording, '0' and '1' bits are written and read by miniature magnetic heads as local magnetization directed 'down' and 'up'. At the same time, such magnetic bit readout can also be performed using the magneto-optical polarization rotation due to the Faraday or Kerr effects, so that $\Phi>0$ corresponds to '1' and $\Phi<0$ -- to '0'. For the conventional magnetic data storage these two ways are identical in the physical sense.

Novel opportunities arise if one creates thickness-modulated magnetic cells and binds the bit value not to the magnetization direction but to the sign of the FR it produces at a certain wavelength. This means that for the correct readout of a bit it becomes necessary to know the wavelength at which the readout should be performed. If the bits are written so that their readout should be made using a certain sequence of the different wavelengths, this sequence serves as an encryption key. It should be noted that without knowing a key the bits can not be read correctly. Fig.~\ref{Fig: Encryption} schematically shows an example of such encryption on three bits of thicknesses 5, 20, and 60~nm all  magnetized upwards. The encryption wavelengths and relevant signs of FR are taken from the experimental data for the samples investigated here. The first and second line pairs wavelength -- FR-sign show that depending on the wavelength keys the bits can be read as any of the possible 8 variants. Moreover, the inverse correspondence is ambiguous: the same bit sequence (3rd line) might be encrypted using several sets of wavelength keys. Thus, the stored data becomes encrypted at the hardware level. Usually, encryption takes place at the software level, so that encrypted data can be directly loaded to some fast processing devices like GPU for decryption. Hardware encryption limits the ability to do this and to brute force the key, as data reading is much slower than operations with RAM and GPU memories.

To conclude, a novel feature of the FR is revealed, which is an ability to observe sign-changing dependence of the FR on the magnetic film thickness. This feature was observed experimentally as the spectral shift of the wavelengths corresponding to the zero FR for the films of the same composition and different thickness. Accurate numerical calculations using the exact analytical equation confirm such a behavior. A simple model considering one-pass propagation of the light through the film is used to reveal the physical mechanism of the sign-changing FR, which is the interplay between its surface and bulk contributions. The necessary condition for observation of the sign-changing FR is relatively large imaginary part of the gyration and refractive index, which means that such an unusual effect takes place in the the absorption band of the material. However, the absorption doesn't represent any obstacle for the Faraday effects measurements since for the nanofilms the transmission remains at high level even for the moderate absorption. Possible application of the observed unusual FR behavior is the hardware data encryption, where the set of light wavelengths at which the magneto-optical bit reading takes place is used as an encryption key. Other potential applications might arise from expanding the abilities for polarization control at nanoscale.

\begin{acknowledgments}
The experimental part of this work was financially supported by the Russian Science Foundation (project No. 24-42-02008). AVB and VIB acknowledge support of the theoretical part of this work by BASIS Foundation (project No. 25-1-1-49-4).
\end{acknowledgments}

\bibliography{_Manuscript}

\end{document}

\title{Unusual sign-changing Faraday effect in nanometer-thick magnetic films\\SUPPLEMENTARY}

\author{Belkova A.V.}
 \email{belkova.av18@physics.msu.ru}
 \affiliation{Faculty of Physics, Lomonosov Moscow State University, Leninskie gori, Moscow 119991, Russia}
\author{Ignatyeva D.O.}
 \affiliation{Faculty of Physics, Lomonosov Moscow State University, Leninskie gori, Moscow 119991, Russia}
\affiliation{Russian Quantum Center, Moscow 121205, Russia}
\author{Kalish A.N.}
 \affiliation{Faculty of Physics, Lomonosov Moscow State University, Leninskie gori, Moscow 119991, Russia}
\affiliation  {Russian Quantum Center, Moscow 121205, Russia}
\author{Vetoshko P.M.}
\affiliation{Institute of Physics and Technology, V.I. Vernadsky Crimean Federal University, Simferopol 295007, Crimea}
\author{Kudryashov A.L.}
\affiliation{Institute of Physics and Technology, V.I. Vernadsky Crimean Federal University, Simferopol 295007, Crimea}
\author{Belotelov V.I.}
\email{belotelov@physics.msu.ru}
\affiliation{Faculty of Physics, Lomonosov Moscow State University, Leninskie gori, Moscow 119991, Russia}
\affiliation{Russian Quantum Center, Moscow 121205, Russia}

\date{\today}

\renewcommand{\theequation}{S.\arabic{equation}}
\renewcommand{\thefigure}{S\arabic{figure}}
\setcounter{figure}{0}  
\setcounter{equation}{0}
%\subsection{Sample Fabrication}
%For the experimental study, a sample of the composition $\mathrm{Bi_1Gd_2Fe_{5}O_{12}}$ grown by liquid-phase epitaxy on a GSGG substrate were chosen. The thickness of the samples varied from 5 to 60 nm. The sample was grown using the following procedure. The substrate for the growth of films is placed in a melt solution. The cylindrical melt tank consists of a chamber made of high-strength stainless steel lined with quartz. The prepared substrates are loaded into the melt through a transfer chamber. The blade assembly can be lowered into the melt and rotated to stir the melt. Typically, high purity melt components are loaded into a pure melt container at room temperature. The temperature in the furnace is raised above the predicted melting point and maintained constant until the entire substance is dissolved. Ferrite garnets are grown from a lead melt solution. Lead oxide is a rather aggressive solvent at temperatures of about 1000 degrees Celsius, so it is necessary to use a platinum crucible. It has sufficient refractoriness and chemical resistance up to 1300 degrees. Usually, a crucible with a preheated charge is placed in a heated oven. Then the desired temperature is set for crystal growth. Then the substrate is lowered from above into the melt. The thickness of the film is adjusted by the time that the substrate spends in the melt.

\maketitle

\subsection{Numerical Simulations}
\begin{figure}[htb!]
\center{\includegraphics[width=0.8\linewidth]{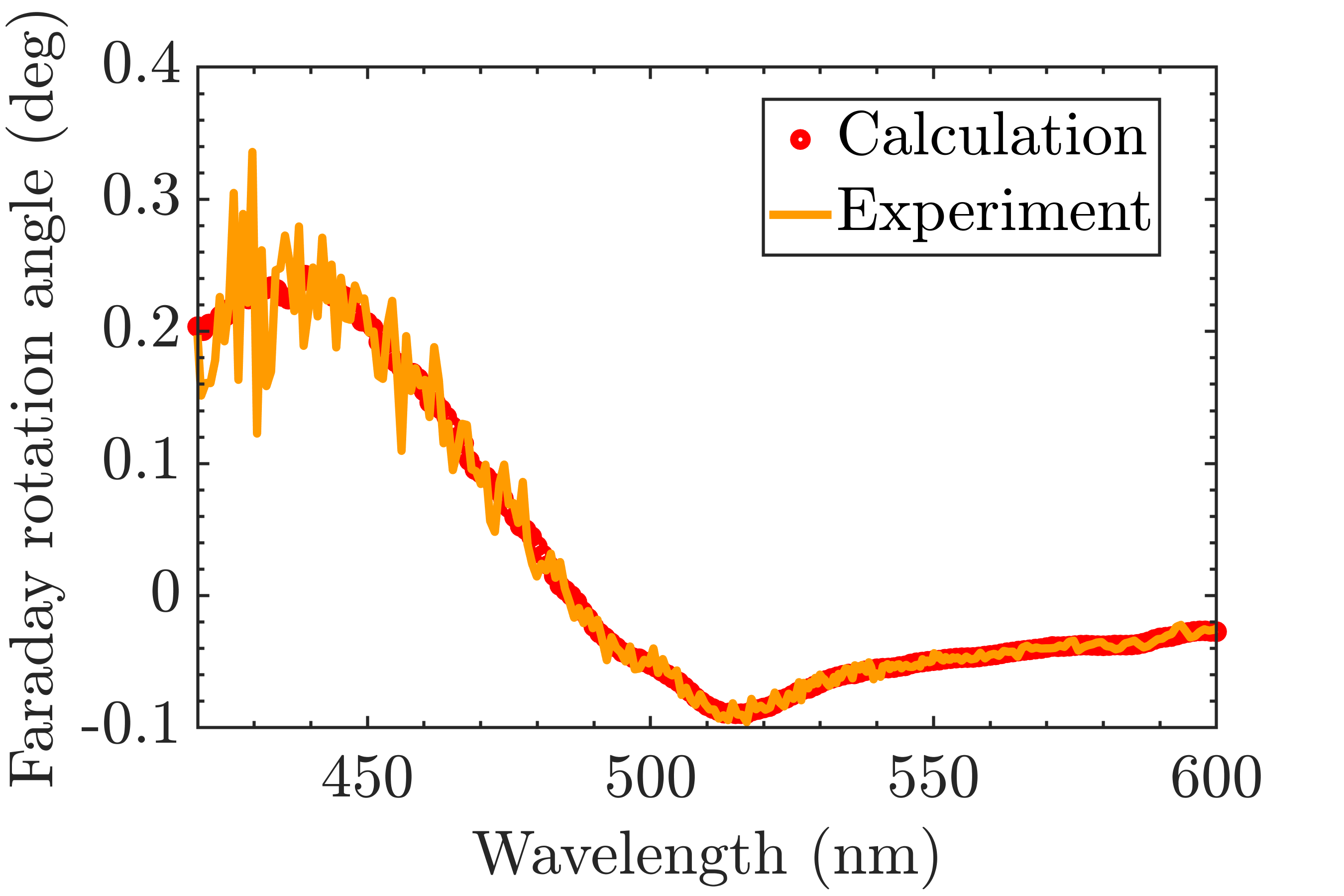}}
\caption{Spectra of the FR angle obtained experimentally for the film of 10~nm thickness (orange curve) and theoretically by Eq.~\eqref{EqC2b} (red circles).}
\label{ris:image4}
\end{figure}
The wavelength dependent diagonal component of the dielectric tensor of the magnetic film $\varepsilon_0(\lambda)$ was taken from ~\cite{wittekoek1975magneto}   and then transformed into $\varepsilon(\lambda) = a\cdot\varepsilon_{0}(b\cdot \lambda + c)$. The transformation coefficients and film thickness were refined for the 10~nm thick film so that the calculated transmission spectrum coincides with the experimental one. The gyration spectrum of this section of the film was found by an iterative procedure. In particular, at the first step, gyration was assumed purely real and $g'(\lambda)$ was found to better match the FR spectra calculated by Eq.~(S.5) and the experimental one. At the next step, the imaginary part of the gyration $g''(\lambda)$ was found by the Kramers-Kronig formula:
 \begin{equation}
     g''(\omega) = -\frac{1}{\pi}v.p.\int_{-\infty}^{+\infty}\frac{g'(\omega_0)\mathrm{d}\omega_0}{\omega_0 - \omega},
\label{eq:5}
 \end{equation}
where $\omega$ is the angular frequency of light.
\begin{figure}[htb]
    \centering
    \includegraphics[width=0.5\textwidth]{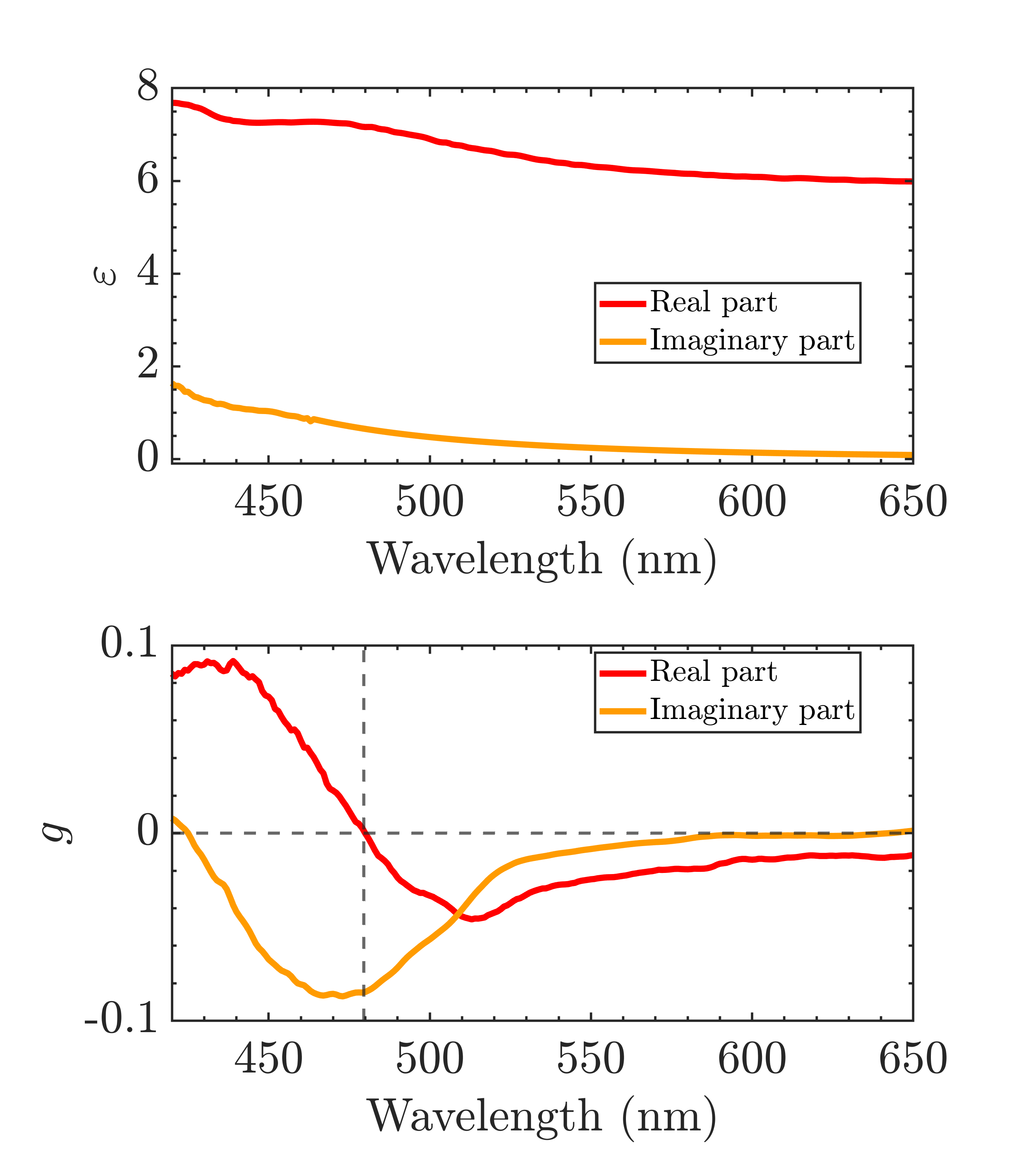} 
    \caption{Obtained parts of the dielectric permittivity tensor for the sample: the diagonal (upper panel) and the non-diagonal (lower panel) components of the dielectric permittivity tensor.}
    \label{fig:perm}
\end{figure}

\begin{figure*}[htb]
\begin{minipage}[h]{0.49\linewidth}
 \begin{flushleft}
    (a)
  \end{flushleft}
\center{
\includegraphics[width=1\linewidth]{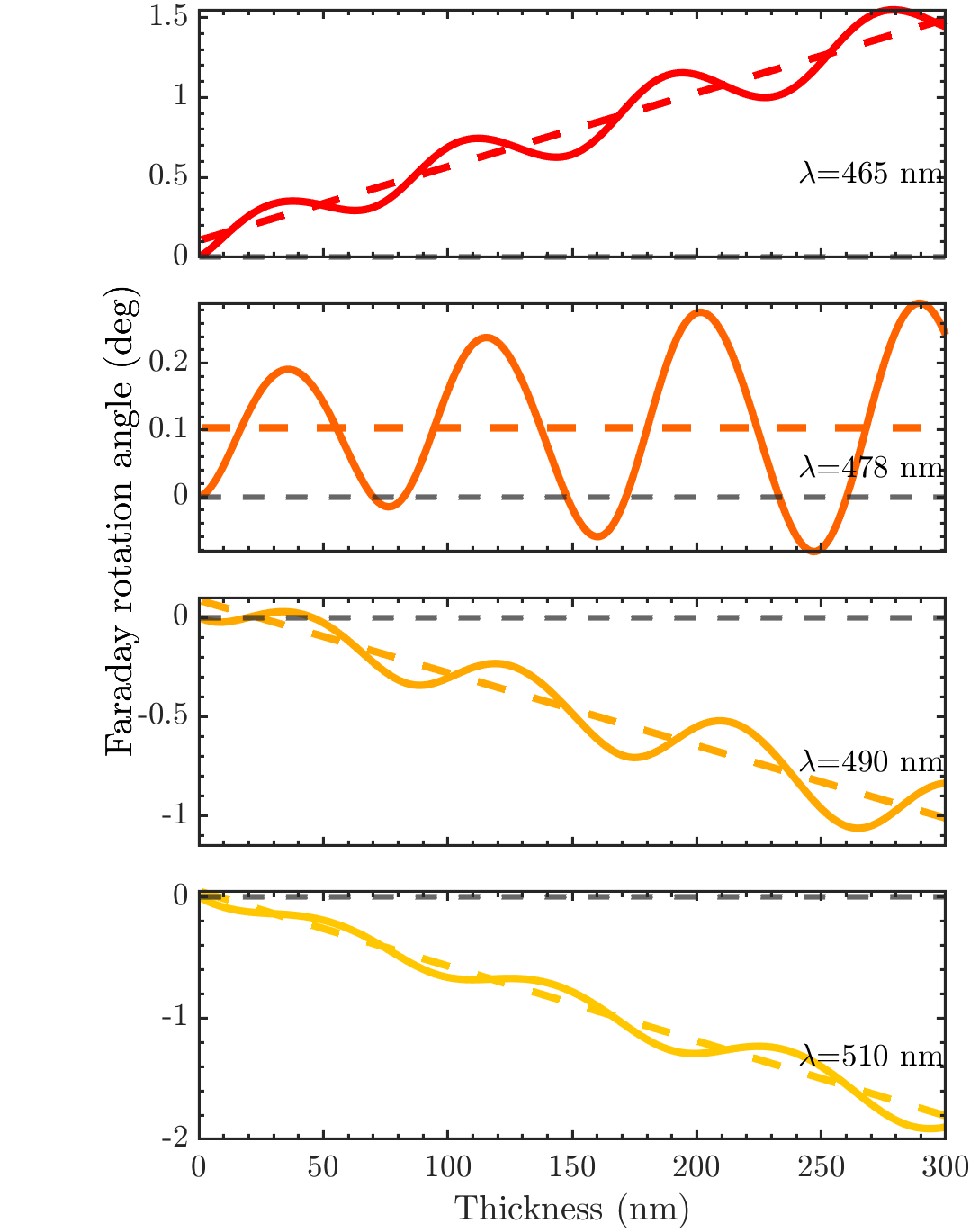}}\\
\end{minipage}
\hfill
\begin{minipage}[h]{0.49\linewidth}
\begin{flushleft}
    (b)
  \end{flushleft}
\center{\includegraphics[width=1\linewidth,height=1.2\linewidth]{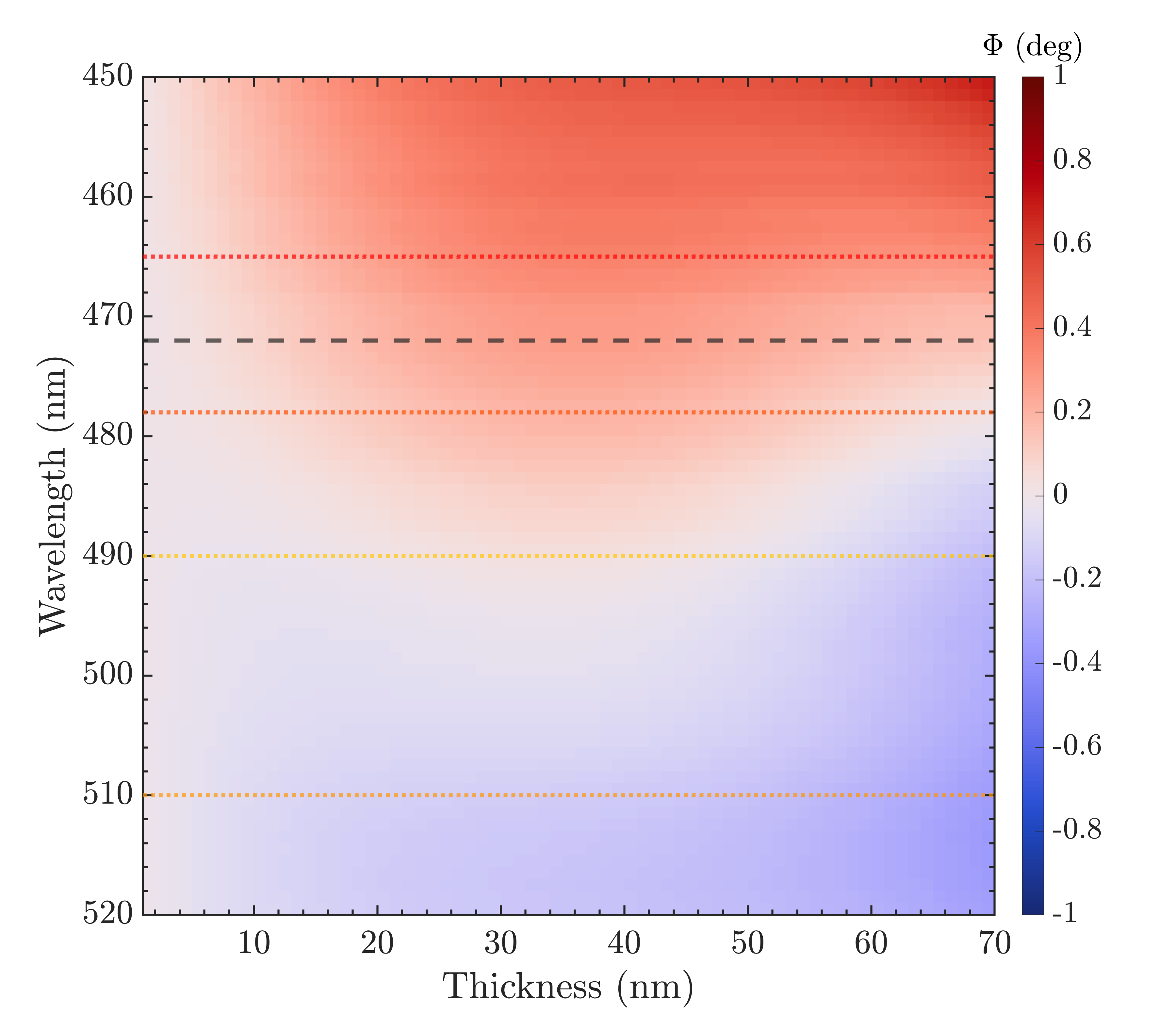}} \\
\end{minipage}
\caption{\label{Fig: General}(a) Theoretically calculated dependencies of the Faraday rotation on the magnetic film thickness by the exact equation Eq.~\eqref{EqC2b} (solid lines) and within the single-pass model (dashed lines). (b) The FR dependence on the wavelength and film thickness. Coloured dotted lines denote the wavelengths $\lambda=465, 478, 490, 510$~nm, black dotted line denotes the wavelength where the FR sign changes according to Eq.~(1).}
\end{figure*}
Further, using the obtained $g''(\lambda)$, a refined $g'(\lambda)$ was found at the next iteration step to better match the FR spectra calculated by Eq.(S.5) and the measured ones, especially for the $\lambda <~$530~nm corresponding to the absorption band. After that the Kramers-Kronig relation was used again to refine  $g''(\lambda)$ and so on. Such iterative process was repeated five times until the calculated FR spectrum almost perfectly coincided with the experimental data (Fig. S1).

Spectra of the obtained components of dielectric permittivity tensor can be seen in Fig.~\ref{fig:perm}. Note that for the found permittivity tensor components the relation $n_m' \gg n_m'' \gg g'\sim g''$ is fulfilled, which justifies analytical considerations presented in the main text. The FR and transmittance spectra for the other magnetic films were calculated based on the found spectra of $n(\lambda)$ and $g(\lambda)$. It allowed to refine values of their thicknesses.

Fig. S3(a) shows the FR dependence on the film thickness obtained via exact analytical calculations.

\subsection{Details on theoretical analysis}
The Faraday rotation is caused by the phase difference between the two circularly polarized waves:
\begin{equation}\label{EqC1}
    \Phi=\frac{1}{2}(\phi_{+}-\phi_{-}),
\end{equation}
where '$+$' and '$-$' denote the two circular polarizations. At the transmission through a film of the thickness $d$ the phase can be determined as an argument of a complex amplitude $t_\pm$ of the transmitted wave, so for the normal incidence:

\begin{equation}\label{EqC2}
\phi_{\pm} =  
\mathrm{arg}\left(t_{\pm}\right), 
\end{equation}

\begin{equation}\label{EqC2a}
t_{\pm} =  
\frac{2n_1n_{2\pm}}
{n_{2\pm}(n_1+n_3)\cos(\alpha_{\pm})
-i(n_{2\pm}^2+n_1n_3)\sin(\alpha_{\pm})}, 
\end{equation}
where $\alpha_{\pm} = k_0n_{2\pm}d$, and indices $1,2,3$ denote the upper medium (air), magnetic material and lower medium (substrate), correspondingly; $n_{2\pm}=\sqrt{n_m^2\pm g}$, and $k_0$ is the vacuum wavenumber. The expression for $t_\pm$ is obtained from the Fresnel equations. Substituting Eqs.~\eqref{EqC2} and~\eqref{EqC2a} to Eq.~\eqref{EqC1}, we come to the exact analytical equation for the FR angle, acquired by light passing through the magnetic film with arbitrary complex $n_m$, $g$ and thickness $d$, placed between the two semi-infinite media with refractive indices $n_1$ and $n_3$:

\begin{equation}\label{EqC2b}
\Phi = \frac{1}{2} \mathrm{arg}\left(\frac{n_{2+}(n_1+n_3)\cos(\alpha_{+})
-i(n_{2+}^2+n_1n_3)\sin(\alpha_{+})}{n_{2-}(n_1+n_3)\cos(\alpha_{-})
-i(n_{2-}^2+n_1n_3)\sin(\alpha_{-})}\right).
\end{equation}

Similarly, we can obtain the separate surface contributions to the phase difference. The phase that the circular polarized wave acquires at the propagation through the upper interface is

\begin{equation}\label{EqC3}
    \phi_{\mathrm{up}\pm}=\mathrm{arg}(t_{12\pm})=\mathrm{arg}\left(\frac{2n_1}{n_1+ n_{2\pm}}\right),
\end{equation}
and for the lower interface

\begin{equation}\label{EqC4}
    \phi_{\mathrm{low}\pm}=\mathrm{arg}(t_{23\pm})=\mathrm{arg}\left(\frac{2n_{2\pm}}{n_{2\pm}+n_3}\right).
\end{equation}

For the further analysis we assume that $n_1$ and $n_3$ are real, and $n_m' \gg n_m'' \gg g'\sim g''$. At this we can apply expansion in $g/n$:

\begin{equation}\label{EqC5}
\begin{split}
    t_{12\pm}=\frac{2n_1}{n_1+n_m}\mp\frac{gn_1}{n_m(n_1+n_m)^2}, \\
    t_{23\pm}=\frac{2n_m}{n_m+n_3}\pm\frac{gn_3}{n_m(n_m+n_3)^2}.
\end{split}
\end{equation}
We can introduce the expansion $t_{ij\pm}=t_{ij}'+it_{ij}''\pm\tau_{ij}'\pm i\tau_{ij}''$, where $\tau_{ij}$ is the contribution of the gyration (i.e., the second terms in Eq.~\eqref{EqC5}). From our assumptions and Eq.~\eqref{EqC5} it follows that $t_{ij}' \gg t_{ij}'' \gg \tau_{ij}\sim \tau_{ij}''$, so

\begin{equation}\label{EqC6}
\begin{split}
    \Delta\phi_{ij}=\phi_{ij+}-\phi_{ij-}=\mathrm{arg}(t_{ij+})-\mathrm{arg}(t_{ij-}) \\ =\tan^{-1}\frac{t_{ij}''+\tau_{ij}''}{t_{ij}'+\tau_{ij}'}-\tan^{-1}\frac{t_{ij}''-\tau_{ij}''}{t_{ij}'-\tau_{ij}'} \\ \approx\left(\frac{t_{ij}''}{t_{ij}'}+\frac{\tau_{ij}''}{t_{ij}'}\right)-\left(\frac{t_{ij}''}{t_{ij}'}-\frac{\tau_{ij}''}{t_{ij}'}\right)=2\frac{\tau_{ij}''}{t_{ij}'}.
\end{split}
\end{equation}

Thus from Eqs.~\eqref{EqC5} and \eqref{EqC6} we obtain

\begin{equation}\label{EqC7}
\begin{split}
    \Delta\phi_{\mathrm{up}}\equiv\Delta\phi_{12}\approx-\frac{g''}{n_m'(n_1+n_m')}, \\
    \Delta\phi_{\mathrm{low}}\equiv\Delta\phi_{23}\approx\frac{g''n_3}{n_m'^2(n_m'+n_3)},
\end{split}
\end{equation}

Finally, from Eqs.~\eqref{EqC7} and~(2) we come to Eq.~(3).

Fig.~S3(a) shows the Faraday rotation obtained within the single-pass model as well as calculated by the exact equation Eq.~\eqref{EqC2b}. One can see that the interference effects that were neglected within the model add nothing but the oscillatory in $d$ behavior.


\subsection{Faraday effect measurements via balanced scheme}
Experimental measurements of the Faraday rotation were carried out using the following scheme. The intensity ($I_+$ and $I_-$, where the $'+'$ sign denotes the forward direction of the field and the $'-'$ sign denotes the backward direction) of broadband light beam from a halogen lamp passing through the sample placed in a constant out-of-plane magnetic field of $\pm200$~mT was measured in the presence of a polarizer and an analyzer crossed at an angle of 45 degrees with each other.

The sample was placed in the magnetic field of 200 mT produced by an electromagnet. After passing through the sample, the light was spectrally analyzed by the spectrometer ASP-IRS AVESTA PROJECT Ltd. Spectra of the transmitted light were measured in the out-of-plane magnetic field for its forward and backward directions: $I_+$ and $I_-$, respectively. Thus, the Faraday rotation angle was found from $\Phi =0.5\arcsin(\frac{I_+ -I_-}{I_+ + I_-})$. 
\begin{figure}[htb!]
\center{\includegraphics[width=0.4\textwidth]{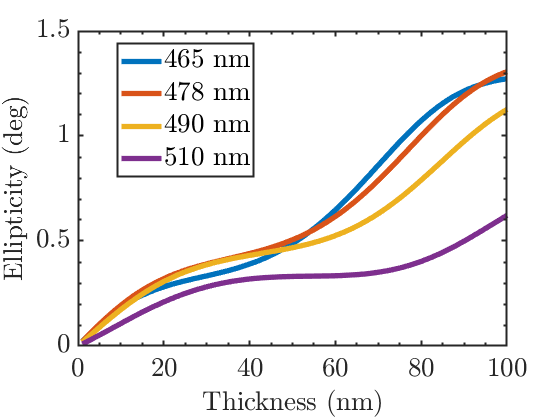}}
\caption{\label{Fig:ellipt} The calculated dependence of ellipticity in the Faraday geometry on film thickness for the wavelengths $\lambda=465, 478, 490, 510$~nm.}
\end{figure}
The paramagnetic contribution from the GSGG substrate was subtracted from the measured spectrum of the Faraday rotation angle.

Let us consider the influence of ellipticity on the measured Faraday rotation angle. The general formula for calculation of the Faraday rotation angle from the measurements in our setup is as follows:

\begin{equation}
\label{eq:faraday}
\begin{aligned}
\Phi &= 0.5\arcsin\left(\frac{I_+ - I_-}{I_+ + I_-}\right) \\
&= 0.5\arcsin\frac{\sin 2\alpha \cdot \sin 2\Phi_0 \cdot (1 + 2\chi)}{1 + \cos 2\alpha \cdot \cos 2\Phi_0 + 4\chi^2 \sin^2\alpha \cos^2\Phi_0}
\end{aligned}
\end{equation}

\noindent where:
$\alpha$ is the angle between the polarizer and the analyzer,
$\Phi_0$ is the Faraday rotation angle to be measured,
$\chi$ is the ellipticity angle.

Therefore, ellipticity might influence the measured FR angle. However, we have checked that in our case (Fig. S4) the ellipticity is small enough so that the measured FR angle $\Phi$ might deviate from the real one $\Phi_0$ by less than 5 percent which is negligible. 
\bibliography{_Manuscript}